\documentclass[letterpaper,12pt]{article}
\usepackage{osajnl}
\usepackage[dvips,colorlinks=true,bookmarks=false,citecolor=blue,
urlcolor=blue,pdfstartview=FitH]{hyperref}

\begin{document}

\title{Integrated Conditional Teleportation and Readout Circuit Based
on a Photonic Crystal Single Chip}

\author{Durdu \"O. G\"uney}
\address{Department of Electrical and Computer Engineering, University
of California, San Diego, 9500 Gilman Dr., La Jolla, California, 92093-0409}
\address{Department of Mathematics, University
of California, San Diego, 9500 Gilman Dr., \\
La Jolla, California, 92093-0112}
\email{dguney@ucsd.edu}

\author{David A. Meyer}
\address{Department of Mathematics, University
of California, San Diego, 9500 Gilman Dr., \\
La Jolla, California, 92093-0112}
\email{dmeyer@math.ucsd.edu}

\begin{abstract}We demonstrate the design of an integrated conditional quantum
teleportation circuit and a readout circuit using a two-dimensional
photonic crystal single chip. Fabrication and testing of the proposed
quantum circuit can be accomplished with current or near future semiconductor
process technology and experimental techniques. The readout part of
our device, which has potential for independent use as an atomic interferometer,
can also be used on its own or integrated with other compatible optical
circuits to achieve atomic state detection. Further improvement of
the device in terms of compactness and robustness could be achieved
by integrating it with sources and detectors in the optical regime.\\
\end{abstract}


\maketitle

\section{Introduction}
It has been shown theoretically by Bennett$^1$, et al.
that quantum teleportation of an unknown state can be achieved using
Bell states. Since then several theoretical and experimental schemes
have been proposed, which include implementation in optical, NMR,
and cavity QED systems$^{2\--5}$. Inspired by the recent paper of S.-B. Zheng$^2$, in this letter we propose a photonic crystal
(PC) based cavity QED implementation of the scheme described therein,
as well as the integration of the device with an atomic readout circuit
to detect the teleported state. Since our proposed device functions
as a very simple conditional teleportation chip, where the success
of the protocol is conditioned on a specific measurement result contrary
to unconditional one, it could be fabricated and tested with current
or near future semiconductor processing technology and experimental
techniques, respectively.

An illustration of the photonic crystal chip that we propose
is shown in Fig.\ 1, where a single mode cavity is integrated with
two parallel waveguides. The actual size of the chip, which is designed
in this letter, is larger than that of the illustration. The actual
height (i.e., vertical) is $\frac{17\sqrt{3}a}{2}$ and the width
(i.e., horizontal) is $79a$, where $a$ is the lattice constant. $x_{1}$, $x_{2}$,
$x_{3}$, $x_{4}$ and $y$ are $9a$, $43a$, $18a$, $9a$, and
$4\sqrt{3}a$, respectively. Because of its large size, we design
the chip in a 2D lattice to reduce the high computational cost of
3D crystal analysis. This is, however, a good approximation due to
the ability of 2D photonic crystal defect modes to emulate those of
3D photonic crystals. This emulation issue is considered in detail
in Ref. {[}6{]} and its advantages for photonic crystal based quantum
logic gate design are discussed in Ref. {[}7{]}, where we have used
exactly the same lattice parameters. Thus, our 2D PC structure is
chosen to emulate its 3D counterpart and is based on a triangular
lattice of silicon rods with dielectric constant of $12$ and lattice
constant of $2.202{\rm mm}$. The radii of the rods and the defects
are $0.175a$ and $0.071a$, respectively. The trajectories of the
atoms through the PC are depicted by the red and blue dashed lines
in Fig.\ 1.

Incorporated with detectors, the cavity part of the chip
operates as a teleportation circuit and the parallel waveguides act
as a readout circuit to detect the teleported atomic
state (see Fig.\ 1). Below we use the terms, teleportation circuit
(TC) and readout circuit (RC), respectively to refer those two integrated
photonic crystal circuits. We choose the width of the former to be
$43a$ and the latter to be $36a$ in order to avoid unwanted coupling
between the two circuits. In the first part of this letter we describe
the teleportation mechanism and in the second part we explain how
to detect the atomic state$^{8\--11}$ within our designed PC quantum
circuit (PCQC).

\section{Teleportation Circuit}
The TC involves the interaction of two Rb atoms with a resonant
high-Q (assumed to be larger than $10^{8}$ to prevent cavity decoherence)
single mode cavity and the detection of one of the atoms using a detector.
The total Hamiltonian (Jaynes-Cummings Model) for the TC under the
dipole and rotating wave approximations is given by$^7$:
\begin{equation}
H_{atom-cavity}=\frac{\hbar\omega}{2}\sum_{j}\sigma_{z}^{j}+\hbar\omega\alpha^{\dagger}\alpha+\hbar\sum_{j}|g(\mathbf{r}_{j})|(\alpha^{\dagger}\sigma_{-}^{j}+\alpha\sigma_{+}^{j})
\end{equation}
where the summation is carried over two atoms, $A$ and
$B$, $\omega$ is the resonant frequency, $\sigma_{z}$ is the $z$-component
of the Pauli spin operator and $\sigma_{\pm}$ are the atomic raising
and lowering operators. $\alpha$ and $\alpha^{\dagger}$ are the
annihilation and creation operators, respectively. $g(\mathbf{r}_{j})$
is the atom-cavity coupling parameter at the position of atom $j$
and can be written as
\begin{equation}
g(\mathbf{r}_{j})=g_{0}E(\mathbf{r}_{j})/|\mathbf{E}(\mathbf{r}_{m})|.
\end{equation}
$g_{0}$ denotes the vacuum Rabi frequency and $\mathbf{r}_{m}$
is the position in the dielectric where the electric field energy
density is maximum. Note that we implicitly assume in Eq. (2)
that the atomic dipole moments, $\mu_{10}^{j}$, of the atoms are
aligned with the photon polarization.

In the first stage of the conditional teleportation protocol,
atom $B$ (whose trajectory is depicted by the red dashed line in
Fig. 1), initially in the excited state $|1\rangle_{B}$, traverses
the cavity, which is initially in the vacuum state $|0\rangle$, with
velocity $V_{B}$. Since at this stage atom $A$, whose state is to
be teleported, is not in play, index $j$ only takes value $B$ in
Eq. (1). The composite state for atom $B$ and the cavity at
time $t$ then can be written as
\begin{equation}
|\Psi(t)\rangle={\rm cos}G_{B}(t)|1\rangle_{B}|0\rangle-i{\rm sin}G_{B}(t)|0\rangle_{B}|1\rangle,\end{equation}
where
\begin{equation}
G_{j}(t)\equiv\int_{t_{0}}^{t}|g(\mathbf{r}_{j})|d\tau.\end{equation}

The cavity we have designed in this letter supports a transverse
magnetic (TM) polarized monopole mode as illustrated in Fig.\ 1.
Based on the block-iterative plane wave expansion method$^{12}$,
with $32$ grid points per $a$, we find the normalized frequency
of the cavity mode to be $0.3733c/a$. By setting $a=2.202{\rm mm}$,
we tune the resonant wavelength to $5.9{\rm mm}$. At this wavelength,
the atomic dipole moment for the Rb atom is taken to be $2\times10^{-26}{\rm Cm}$.$^7$

Using these parameters, we obtain $G_{B}(t)\cong9\pi/4$
at time $t_{1}=51.6{\rm \mu s}$ at $18a$ from the left edge of the
TC, if $V_{B}$ is set to $767.7{\rm m/s}$, and hence atom-cavity
entanglement in Eq. (3). As atom B travels across the cavity,
the time-dependent atom-cavity coupling parameter and the evolution
of the probability amplitudes of states $|1\rangle_{B}|0\rangle$
(blue) and $-i|0\rangle_{B}|1\rangle$ (pink) in Eq. (3) are
displayed in Fig.\ 2. In our calculations we neglect the insignificant
effect of the mode tail on the probability amplitudes of the entangled
state---less than $1\%$ beyond $18a$. Another even less significant
source of error in our design is the coupling of the waveguides in
the RC to the cavity$^{13,14}$, which is simply circumvented by
keeping $x_{2}$ in Fig.\ 1 sufficiently large (i.e., $43a$).

Having entangled atom $B$ and the cavity, the second stage
of the conditional teleportation protocol is to inject atom $A$ (whose
trajectory is illustrated by the blue dashed line in Fig.\ 1), which
is initially (i.e., at time $t_{1}$) in an arbitrary (or an unknown)
quantum state,
\begin{equation}
|\phi\rangle_{A}=c_{0}|0\rangle_{A}+c_{1}|1\rangle_{A}.\end{equation}
Thus at time $t_{1}$, the state of the whole system is
written as
\begin{equation}
|\varphi(t_{1})\rangle\cong\frac{1}{\sqrt{2}}(c_{0}|0\rangle_{A}+c_{1}|1\rangle_{A})(|1\rangle_{B}|0\rangle-i|0\rangle_{B}|1\rangle).\end{equation}

Once atom $A$ arrives at the detector, right before the
detection, say at time $t_{1}+t_{2}$, the state of the whole has
evolved into
\begin{eqnarray}
|\varphi(t_{1}+t_{2})\rangle & \cong & \frac{1}{\sqrt{2}}\{
c_{0}|0\rangle_{A}|1\rangle_{B}|0\rangle-ic_{0}|0\rangle_{B}[{\rm
cos}G_{A}(t_{2})|0\rangle_{A}|1\rangle-i{\rm
sin}G_{A}(t_{2})|1\rangle_{A}|0\rangle]\nonumber \\
& & +c_{1}|1\rangle_{B}[{\rm cos}G_{A}(t_{2})|1\rangle_{A}|0\rangle-i{\rm
sin}G_{A}(t_{2})|0\rangle_{A}|1\rangle]\nonumber \\
& & -ic_{1}|0\rangle_{B}[{\rm cos\sqrt{2}}G_{A}(t_{2})|1\rangle_{A}|1\rangle-i{\rm
sin\sqrt{2}}G_{A}(t_{2})|0\rangle_{A}|2\rangle]\},
\end{eqnarray}
which would reduce to Eq. (6) of Ref. (2), if $|g(\mathbf{r}_{j})|$
were spatially uniform in Eq. (1).

In the derivation of Eq. (7), the contribution of $|g(\mathbf{r}_{B})|$ in Eq. (1) is neglected, since atom $B$ is sufficiently far from the cavity when atom $A$ is injected. At times close to $t_{1}$, although $|g(\mathbf{r}_{B})|\approx|g(\mathbf{r}_{A})|$, neither atom alone nor combined can start the Rabi oscillation.$^7$ On the other hand, as atom $A$ proceeds its coupling to the cavity mode increases, while the coupling of atom $B$ decreases. This means that the effect of $|g(\mathbf{r}_{B})|$ becomes even less significant for times greater than $t_{1}$.  

The third stage of the conditional teleportation protocol
is the measurement of the state of atom $A$. If the result is $|1\rangle_{A}$,
the combined state of atom $B$ and the cavity becomes
\begin{equation}
\Psi(t_{1}+t_{2})\cong\frac{-c_{0}{\rm sin}G_{A}(t_{2})|0\rangle_{B}|0\rangle+c_{1}{\rm cos}G_{A}(t_{2})|1\rangle_{B}|0\rangle-ic_{1}{\rm cos}\sqrt{2}G_{A}(t_{2})|0\rangle_{B}|1\rangle}{[|c_{0}|^{2}{\rm sin^{2}}G_{A}(t_{2})+|c_{1}|^{2}{\rm cos^{2}}G_{A}(t_{2})+|c_{1}|^{2}{\rm cos^{2}}\sqrt{2}G_{A}(t_{2})]^{1/2}}.\end{equation}
The denominator in Eq. (8) is simply the norm of the
unnormalized state in the numerator. We obtain $G_{A}(t_{2})\cong7\pi/4$
in Eq. (8) if we set $V_{A}=987{\rm m/s}$, and thus
\begin{equation}
|\phi\rangle_{B}\cong c_{0}|0\rangle_{B}+c_{1}|1\rangle_{B}.\end{equation}
That is, the initial arbitrary (or unknown) state of atom
$A$ has been teleported to atom $B$, conditional on detecting atom
$A$ in state $|1\rangle_{A}$, which occurs with probability $1/4$.
The time-dependent coupling parameter for atom $A$ and the evolution
of the probability amplitudes of states $-c_{0}|0\rangle_{B}|0\rangle$
(shown in pink), $c_{1}|1\rangle_{B}|0\rangle$ (shown in blue), and
$-ic_{1}|0\rangle_{B}|1\rangle$ (shown in yellow) for the unnormalized
state (i.e., the numerator) in Eq. (8) are shown in Fig.\ 3.
Note that as time elapses Eq. (8) transforms into Eq. (9).
Note also that atom $B$ arrives at $31a$ at time $t_{1}+t_{2}$
(i.e., $88.9{\rm \mu s}$). After an additional distance of $12a$,
the fourth stage of our conditional teleportation and detection protocol
begins as atom $B$ enters the RC at time $t_{1}+t_{2}+t_{3}$ (i.e.,
$123.3{\rm \mu s}$). Next we study how to detect its state.

\section{Readout Circuit}
We can write the atom-maser Hamiltonian, under the dipole
and rotating wave approximations, as$^{15}$
\begin{equation}
H_{atom-maser}=\frac{\hbar\omega}{2}\sigma_{z}+\hbar\frac{\Omega(\mathbf{r})}{2}(e^{i\omega_{m}t}\sigma_{-}+e^{-i\omega_{m}t}\sigma_{+})\end{equation}
where
\begin{equation}
\Omega(\mathbf{r})=\frac{2\mu_{10}|\mathbf{E}(\mathbf{r})|}{\hbar}\end{equation}
is the Rabi frequency and $\omega_{m}$ is the frequency
of the maser, whose polarization is also matched with the atomic dipole
moment, $\mathbf{\mu}_{10}$, of the atom. In the following we will
first assume a constant $\Omega(\mathbf{r})=\Omega_{0}$ to simplify
the explanation of the detection mechanism, and then we will modify
the analysis to apply to our proposed PCQC, where $\Omega(\mathbf{r})$
is not constant.

Assume that atom $B$ in state $|\phi\rangle_{B}$ {[}see
Eq. (9){]} enters a uniform field region at time $t$ and interacts
during a time interval $t_{4}$. Then its probability amplitudes at
time $t+t_{4}$ can be computed to be$^{16}$
\begin{equation}
c_{0}(t+t_{4})=\{[i\cos\Theta\sin(\frac{\Lambda t_{4}}{2})+\cos(\frac{\Lambda t_{4}}{2})]c_{0}+i\sin\Theta\sin(\frac{\Lambda t_{4}}{2})e^{i\omega_{m}t}c_{1}\} e^{i\omega_{m}t_{4}/2},\end{equation}
\begin{equation}
c_{1}(t+t_{4})=\{ i\sin\Theta\sin(\frac{\Lambda t_{4}}{2})e^{-i\omega_{m}t}c_{0}+[-i\cos\Theta\sin(\frac{\Lambda t_{4}}{2})+\cos(\frac{\Lambda t_{4}}{2})]c_{1}\} e^{-i\omega_{m}t_{4}/2}.\end{equation}
where ${\rm cos}\Theta$, ${\rm sin}\Theta$, and $\Lambda$
are defined as follows.
\begin{equation}
\Lambda\equiv\sqrt{(\omega-\omega_{m})^{2}+\Omega_{0}^{2}},\end{equation}
\begin{equation}
\cos\Theta\equiv\frac{(\omega-\omega_{m})}{\Lambda},\end{equation}
\begin{equation}
\sin\Theta\equiv-\frac{\Omega_{0}}{\Lambda}.\end{equation}

After this interaction, atom $B$ interacts with a second
uniform field region for the same time interval $t_{4}$ starting
at time $t+t_{4}$. The probability amplitudes, $c_{0}(t+2t_{4})$
and $c_{1}(t+2t_{4})$, in this case can be also found from equations
(12) and (13) by replacing $c_{0}$, $c_{1}$, and $t$ with $c_{0}(t+t_{4})$,
$c_{1}(t+t_{4})$, and $t+t_{4}$, respectively.

Note that one can also treat these two uniform field regions
as a single uniform field region with an interaction time of $2t_{4}$.
In this letter, however, we also intend to show that our RC has the
potential to be an atomic interferometer, like a Ramsey interferometer.$^{16}$ In order to hint how this could work, we design two separate
oscillating field regions based on two parallel waveguides as discussed
below. Implementation of a fully working interferometer, however,
requires further consideration.

After atom $B$ has interacted with the two uniform field
regions, we can implicitly write the final state of atom $B$ at time
$t+2t_{4}$, before detection as
\begin{equation}
|\phi(t+2t_{4})\rangle_{B}\cong(c_{0}c_{00}+c_{1}c_{10})|0\rangle_{B}+(c_{0}c_{01}+c_{1}c_{11})|1\rangle_{B}\end{equation}
where $c_{ij}$ is the probability amplitude of the atom
being in state $|j\rangle$ at time $t+t_{4}$ if it is initially
prepared in state $|i\rangle$. For example if the atom is initially
in state $|1\rangle$ then it evolves into $c_{10}|0\rangle+c_{11}|1\rangle$
at time $t+t_{4}$.

Having interacted with both waveguides, atom $B$ is finally
detected by an ionization detector, as shown in Fig.\ 1. The probability
of finding it in the excited state is
\begin{equation}
P_{1}=|c_{0}|^{2}|c_{01}|^{2}+|c_{1}|^{2}|c_{11}|^{2}+[c_{0}^{(r)}c_{1}^{(r)}+c_{0}^{(i)}c_{1}^{(i)}]2(c_{01}c_{11}^{\star})^{(r)}+[c_{0}^{(r)}c_{1}^{(i)}+c_{0}^{(i)}c_{1}^{(r)}]2(c_{01}c_{11}^{\star})^{(i)}.\end{equation}
Superscripts $(r)$ and $(i)$ represent the real and imaginary
part, respectively.

We can simplify Eq. (18) by writing Eq. (9) in the form
\begin{equation}
|\phi\rangle_{B}\cong\cos(\theta/2)|0\rangle_{B}+\sin(\theta/2)e^{i\varphi}|1\rangle_{B}\end{equation}
where we have implicitly set $c_{0}^{(r)}=\cos(\theta/2)$,
$c_{0}^{(i)}=0$, $c_{1}^{(r)}=\sin(\theta/2)\cos\varphi$, and $c_{1}^{(i)}=\sin(\theta/2)\sin\varphi$.

Then, at four different $\Delta\equiv\omega_{m}-\omega$ values we
evaluate $|c_{01}|^{2}$, $|c_{11}|^{2}$, $2(c_{01}c_{11}^{\star})^{(r)}$,
and $2(c_{01}c_{11}^{\star})^{(i)}$ directly and measure $P_{1}$.
Thus having obtained four equations, we can find $|c_{0}|^{2}$, $|c_{1}|^{2}$,
$[c_{0}^{(r)}c_{1}^{(r)}+c_{0}^{(i)}c_{1}^{(i)}]$, $[c_{0}^{(r)}c_{1}^{(i)}-c_{0}^{(i)}c_{1}^{(r)}]$ and hence $|\phi\rangle_{B}$. Note that, although we have only two
unknowns in Eq. (19), it is convenient to use four equations due
to the form of Eq. (18). This demonstrates that one could detect
the teleported state with our proposed device. In order to measure
$P_{1}$, we need an atomic beam of identically prepared single atoms,
since multiple observations are necessary to estimate the probability.
Since the TC only has a 25 percent conditional success rate, incoming
atoms to the RC will not be identical. Thus if atom $A$ of the relevant
pair is not detected in the excited state, the corresponding atom
$B$ measurement is discarded from the calculation of $P_{1}$.

Finally, we describe how to apply these ideas to our proposed
PCQC. Remember that atom $B$ in state $|\phi\rangle_{B}$ enters
the RC at time $t_{1}+t_{2}+t_{3}$. The steady-state mode profiles
of the waveguides are calculated by two-dimensional-finite-difference-time-domain
(2D-FDTD) method with discretizations of $\frac{a}{13}$ and $\frac{a}{13}\frac{\sqrt{3}}{2}$
in the horizontal and vertical directions, respectively. We have designed
the waveguides as coupled-cavity-waveguides$^{17,18}$ to allow the
atoms sufficiently large void regions through which to travel freely
without a resonant-dipole-dipole-interaction or the Casimir-Polder
effect (see Fig.\ 1). The steady-state cross section of the electric
field magnitude, for $\Delta=0$, along the path of atom $B$ is shown
in Fig.\ 4. At the steady state the coupling between the two parallel
waveguides$^{19\--22}$ is not significant.

In our calculations we assume that atom $B$ interacts with
two separate nonuniform waveguide modes of total width $36a$, shown
by the pink curve in Fig.\ 4. In other words we truncated their tails
(shown by blue in Fig.\ 4.), which overlap with the cavity mode.
The effect of this truncation in our design is observed to be negligibly
small. We have observed that the coupling of waveguide modes on the
cavity mode is less than the cavity coupling to waveguides, which
does not effect the probability amplitudes more than $0.6\%$.

Since there is no feedback mechanism, as there is in the
atom-cavity interaction, the atom-waveguide interaction can be described
by a Markoff approximation. In other words, when an atom emits a photon
into the waveguide, it leaves the interaction region immediately and
cannot act back on the atom. Then the waveguides must be designed
in such a way that total rate of spontaneous emission into the waveguide
and the free-space lossy modes is sufficiently low as atom $B$ traverses
them. In our design considering estimated effective transverse cross
section of corresponding three-dimensional waveguide mode the maximum
probability of spontaneous emission for a given atom is estimated
to be on the order of $10^{-4}$.$^{23}$ Thus, it is safe to assume that the Bloch vector of atom $B$ can be manipulated without being correlated with the defect mode$^{24}$
of the waveguides.

The total interaction time with the parallel waveguides
is also $2t_{4}$, as in uniform case above, with $t_{4}$ (i.e.,
$51.6{\rm \mu s}$) for each. Note that, since the Rabi frequency,
$\Omega(\mathbf{r})$, is proportional to $|\mathbf{E}(\mathbf{r})|$,
the pink part of the curve in Fig.\ 4 also describes the position
dependence of the Rabi frequency if multiplied by $2\mu_{10}/\hbar$.
Thus the evolution of the teleported state, as it interacts with the
parallel waveguides, can be determined numerically by exploiting equations
(12) and (13) successively, $t_{4}$ replaced with the maximum tolerated
time-step, which is $44{\rm ns}$ in our case (i.e. time for atom
$B$ to travel $a/65$).

Thus, right before the detection, the final state is $|\phi(t_{1}+t_{2}+t_{3}+2t_{4})\rangle_{B}$,
which can be written exactly in the form of Eq. (17). The probability
of detecting atom $B$ in the excited state is then given by Eq.
(18), and following the same method as in the uniform case above,
we can easily detect $|\phi\rangle_{B}$, the state output by TC.

As an example, assume that the final state of the TC [see
Eq. (19)] is\begin{equation}
|\phi\rangle_{B}={\rm cos}(\pi/8)|0\rangle_{B}+{\rm sin}(\pi/8)e^{-i\pi/6}|1\rangle_{B}\end{equation}

The corresponding calculated values of $|c_{01}|^{2}$ (green),
$|c_{11}|^{2}$ (red), $2(c_{01}c_{11}^{\star})^{(r)}$ (light blue),
$2(c_{01}c_{11}^{\star})^{(i)}$ (purple), and the value of $P_{1}$
(dark blue), which will be measured, are shown in Fig.\ 5. Picking
four different values of $\Delta$, we can easily determine that $\theta=\pi/4$
and $\varphi=-\pi/6$.

\section{Discussions}
Due to the electromagnetic scalability of photonic crystals
one could also implement this device in more compact form in the optical
regime. In this case, however, the electric field gradient and the
momentum recoil of the atom would also have to be analyzed. In the
microwave regime studied in this letter, the small momentum recoil
due to a microwave photon does not cause external degrees of freedom
(i.e., the position or momentum) of the atom to couple with its internal
degrees of freedom (i.e., the ground or excited state).

To implement our PCQC, one should also engineer the surface
states, which reside at the air-photonic crystal interfaces. These
surface modes may also become correlated with the atoms exiting the
crystal. These modes can be handled, however, by appropriate termination
of the crystal.$^{25\--28}$

\section{Conclusions}
In conclusion, we have designed an integrated conditional
teleportation circuit with readout in a single PC chip. Our proposed
PCQC can be implemented by current or near future semiconductor processing
technologies and experimental techniques. Even the detectors and sources
illustrated in Fig.\ 1 could also be integrated---the feasibility
of single atom detection on a PC chip has already been demonstrated$^{29}$, as have PC based lasers.$^{30,31}$ The RC part of our
device can not only be used with the TC but also can be used independently,
or integrated with other compatible optical circuits to read out atomic
states. It also hints at the potential for implementation of atomic
interferometers, such as the Ramsey interferometer, in photonic crystals.

Achieving good mode-matching stability becomes quite cumbersome
for many multiply nested interferometers in optical quantum computing.
Photonic crystals, however, may be an especially promising paradigm
as robust quantum circuit boards for delicate, next generation, scalable
optical quantum computing and networking technologies, as well as
scalable quantum dynamic random access memory$^{32}$, where qubits
are refreshed using PC based cyclical quantum circuits. Integrated
PC devices, such as planar waveguides and elementary interferometer
modules, could replace the current bulky elements in linear optical
quantum circuits. Furthermore, such devices could be made quite compact
using PC integrated with single photon sources.$^{33}$ 

\section*{Acknowledgments}
This work was supported in part by the National Security 
Agency (NSA) and Advanced Research and Development Activity (ARDA) 
under Army Research Office (ARO) Grant No.\ DAAD19-01-1-0520, by 
the DARPA QuIST program under contract F49620-02-C-0010, and by
the National Science Foundation (NSF) under grant ECS-0202087.

\newpage
\section*{List of Figure Captions}

Fig. 1. Illustration of the proposed photonic crystal quantum circuit. This illustration is not to scale.

\noindent Fig. 2. (a) The time-dependent atom-cavity coupling parameter and (b) the evolution of the probability amplitudes of states $|1\rangle_{B}|0\rangle$ (blue) and $-i|0\rangle_{B}|1\rangle$ (pink) in Eq. (3), as atom $B$ travels across the cavity.

\noindent Fig. 3. (a) Time-dependent coupling parameter for atom $A$ and (b) the evolution of the probability amplitudes of states $-c_{0}|0\rangle_{B}|0\rangle$ (pink), $c_{1}|1\rangle_{B}|0\rangle$ (blue), and $-ic_{1}|0\rangle_{B}|1\rangle$ (yellow) for the unnormalized state (i.e., numerator) in Eq. (8).

\noindent Fig. 4. The steady-state cross section of the electric field magnitude, for $\Delta=0$, along the path of atom $B$ (see text).

\noindent Fig. 5. Calculated values of $|c_{01}|^{2}$ (green), $|c_{11}|^{2}$ (red), $2(c_{01}c_{11}^{\star})^{(r)}$ (light blue), $2(c_{01}c_{11}^{\star})^{(i)}$ (purple), and to be measured value of $P_{1}$ (dark blue) (see text).

\newpage

  \begin{figure}[htbp]
  \setlength{\abovecaptionskip}{2.5cm}
  \centering
  \includegraphics[width=14.6cm]{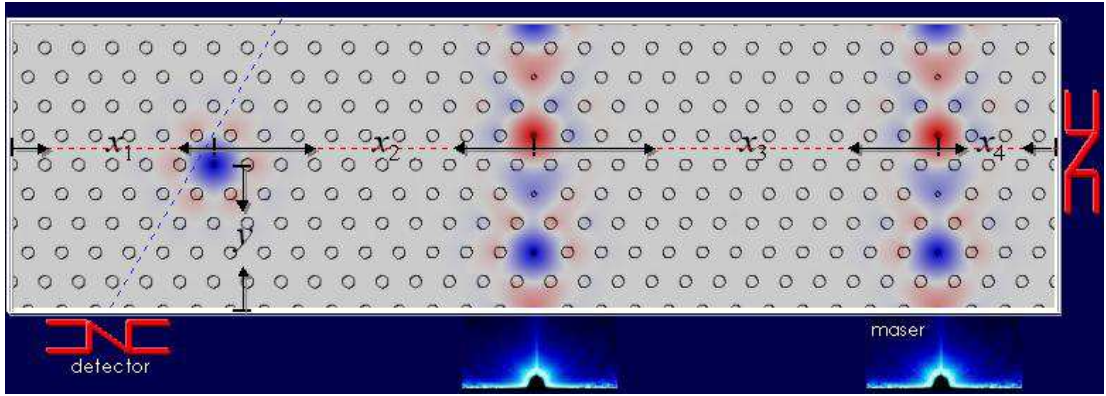}
  \caption{Illustration of the proposed photonic crystal quantum circuit. This illustration is not to scale. guneyF1.eps.}
  \end{figure}

\newpage

  \begin{figure}[htbp]
  \setlength{\abovecaptionskip}{2.5cm}
  \centering
  \includegraphics[width=8.3cm]{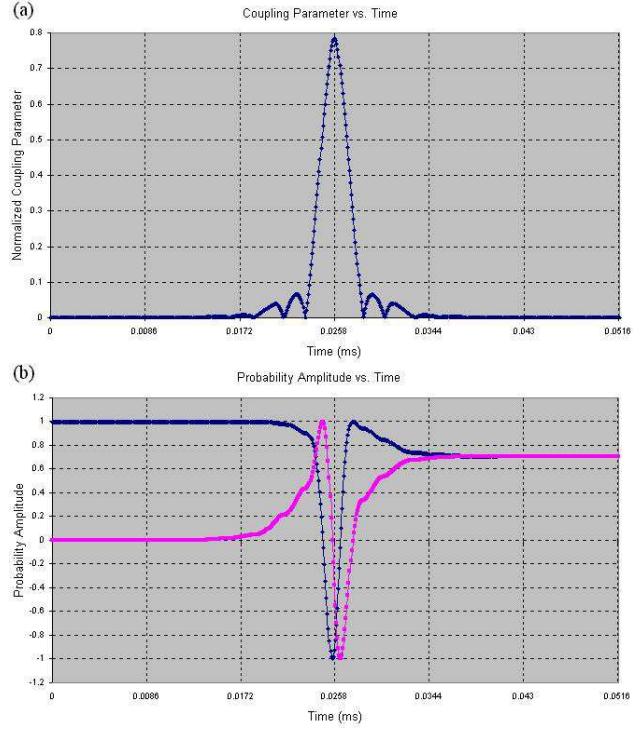}
  \caption{(a) The time-dependent atom-cavity coupling parameter and (b) the evolution of the probability amplitudes of states $|1\rangle_{B}|0\rangle$ (blue) and $-i|0\rangle_{B}|1\rangle$ (pink) in Eq. (3), as atom $B$ travels across the cavity. guneyF2.eps.}
  \end{figure}

\newpage

  \begin{figure}[htbp]
  \setlength{\abovecaptionskip}{2.5cm}
  \centering
  \includegraphics[width=8.3cm]{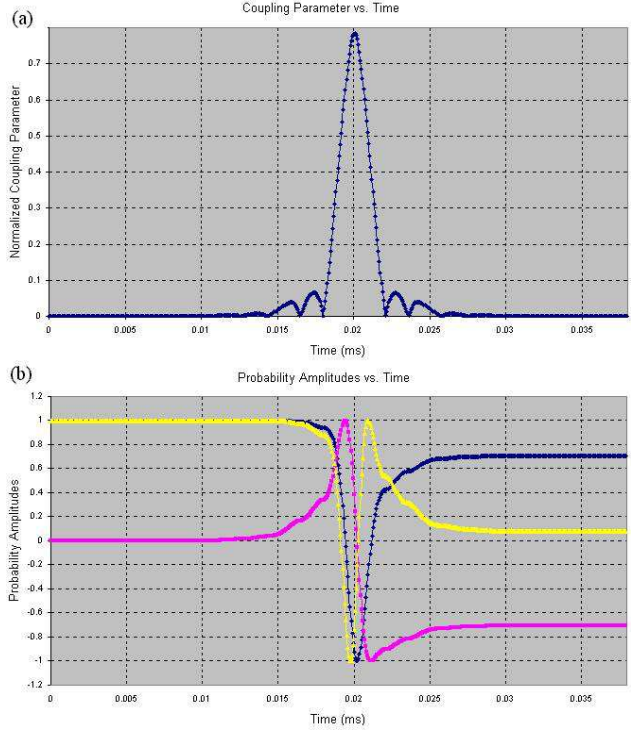}
  \caption{(a) Time-dependent coupling parameter for atom $A$ and (b) the evolution of the probability amplitudes of states $-c_{0}|0\rangle_{B}|0\rangle$ (pink), $c_{1}|1\rangle_{B}|0\rangle$ (blue), and $-ic_{1}|0\rangle_{B}|1\rangle$ (yellow) for the unnormalized state (i.e., numerator) in Eq. (8). guneyF3.eps.}
  \end{figure}

\newpage

  \begin{figure}[htbp]
  \setlength{\abovecaptionskip}{2.5cm}
  \centering
  \includegraphics[width=8.3cm]{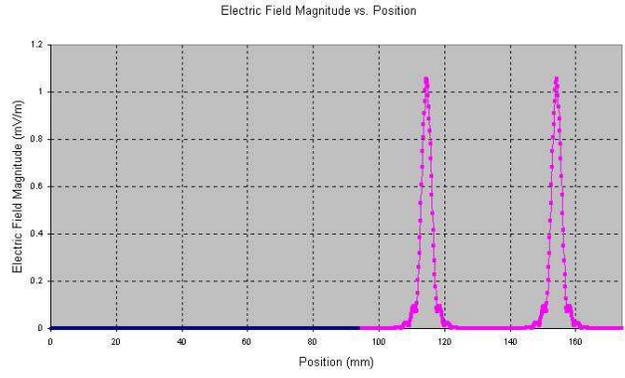}
  \caption{The steady-state cross section of the electric field magnitude, for $\Delta=0$, along the path of atom $B$ (see text). guneyF4.eps.}
  \end{figure}

\newpage

  \begin{figure}[htbp]
  \setlength{\abovecaptionskip}{2.5cm}
  \centering
  \includegraphics[width=8.3cm]{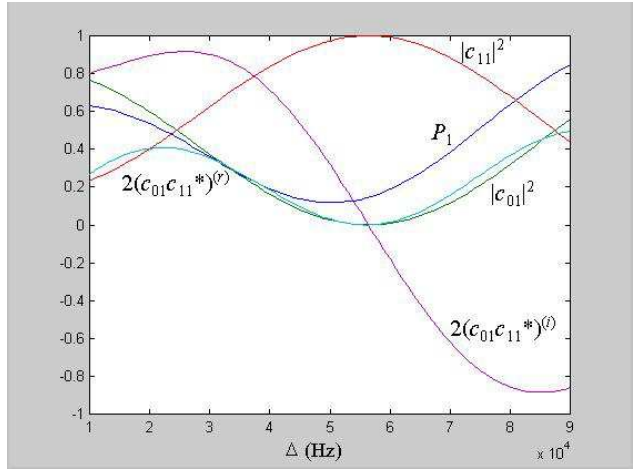}
  \caption{Calculated values of $|c_{01}|^{2}$ (green), $|c_{11}|^{2}$ (red), $2(c_{01}c_{11}^{\star})^{(r)}$ (light blue), $2(c_{01}c_{11}^{\star})^{(i)}$ (purple), and to be measured value of $P_{1}$ (dark blue) (see text). guneyF5.eps.}
  \end{figure}

\end{document}